# Electrochemical etching strategy for shaping monolithic 3D structures from 4H-SiC wafers

*André Hochreiter, Fabian Groß, Morris-Niklas Möller, Michael Krieger, Heiko B. Weber\**

*heiko.weber@fau.de

Department of Physics, Friedrich-Alexander-Universität Erlangen-Nürnberg (FAU), Erlangen D-91058, Germany

Abstract:
Silicon Carbide (SiC) is an outstanding material, not only for electronic applications, but also for projected functionalities in the realm of photonic quantum technologies, nano-mechanical resonators and photonics on-a-chip. For shaping 3D structures out of SiC wafers, predominantly dry-etching techniques are used. SiC is nearly inert with respect to wet-etching, occasionally photoelectrochemical etching strategies have been applied. Here, we propose an electrochemical etching strategy that solely relies on defining etchable volumina by implantation of p-dopands. Together with the inertness of the n-doped regions, very sharp etching contrasts can be achieved. We present devices as different as monolithic cantilevers, disk-shaped optical resonators and membranes etched out of a single crystal wafer. The high quality of the resulting surfaces can even be enhanced by thermal treatment, with shape-stable devices up to and even beyond 1550°C. The versatility of our approach paves the way for new functionalities on SiC as high-performance multi-functional wafer platform.

INTRODUCTION: Silicon Carbide (SiC), especially its polytype 4H-SiC, is an extraordinary material for integrating electronics [1], photonics [2], high-quality mechanics [3] and quantum technologies on the very same chip [4, 5]. Due to its technological breakthrough in power electronics it is available as single crystalline high-quality wafers. When, further, optical and mechanical functionality is demanded, there is a need for highest quality devices with three-dimensional geometries. As to optics, SiC provides the unusual opportunity of simultaneous $\chi^{(2)}$ and $\chi^{(3)}$ nonlinearities [6, 7]. The current state-of-the-art photonics on-a-chip is not integrated with traditional SiC fabrication techniques, but uses thin SiC-on-Insulator technology [4, 8, 9]. As to mechanics on-a-chip, SiC provides an outstanding intrinsic property: it has the lowest internal damping of all known materials [9, 3, 10]. Also here, device fabrication utilizes thin Silicon Carbide layers on a sacrificial substrate. Given the extraordinary set of parameters of SiC, it is desirable to identify *monolithic* technologies for the preparation of optical and mechanical devices along with the electronic functionality. Compatibility with high temperature protocols, for example epitaxial graphene growth [11] or defect annealing [12], would be beneficial.

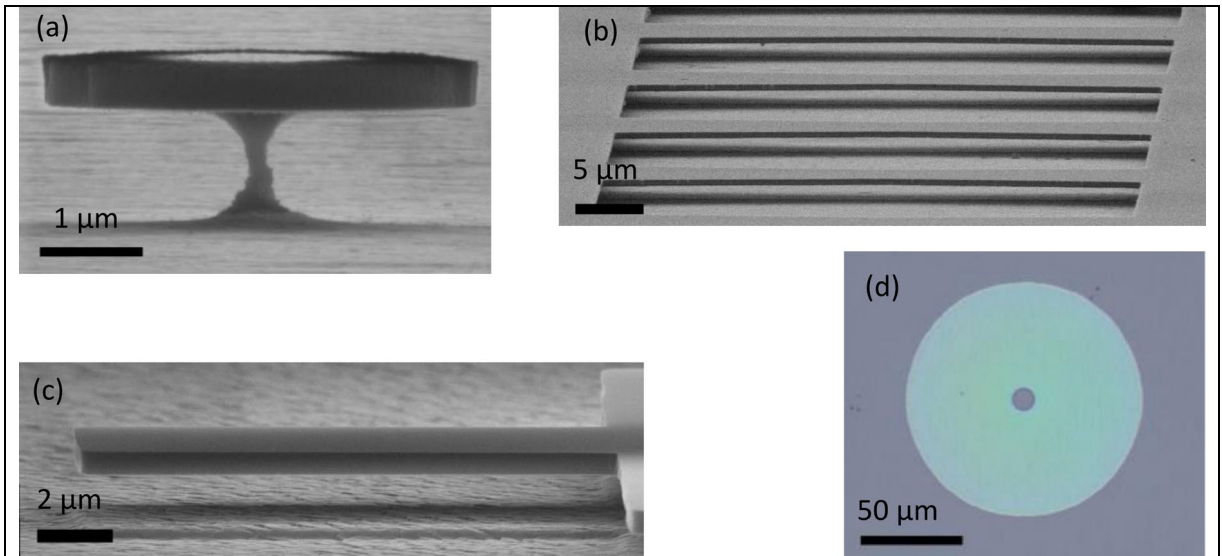

**Figure 1: Monolithically etched 3D-devices from single-crystal 4H-SiC wafer.** (a) disk-shape optical resonator, (b) doubly clamped mechanical resonator, (c) single clamped mechanical resonator, (d) free-standing circular membrane (central hole is required as etching access). (a) – (c) SEM micrographs, (d) optical micrograph.

For such applications, however, a technological prerequisite is an etching strategy that forms the desired 3D-structures monolithically out of the single-crystal wafer, while maintaining high quality surfaces and low defect budgets. The commonly used gas etching strategies (ICP-RIE/RIE) are projective and have limitations, with respect to possible 3D geometries. Further, they are prone to create surface-near point-defects. Therefore, a new process strategy is required. Here, we present a route based on implantation and subsequent electrochemical etching (ECE).

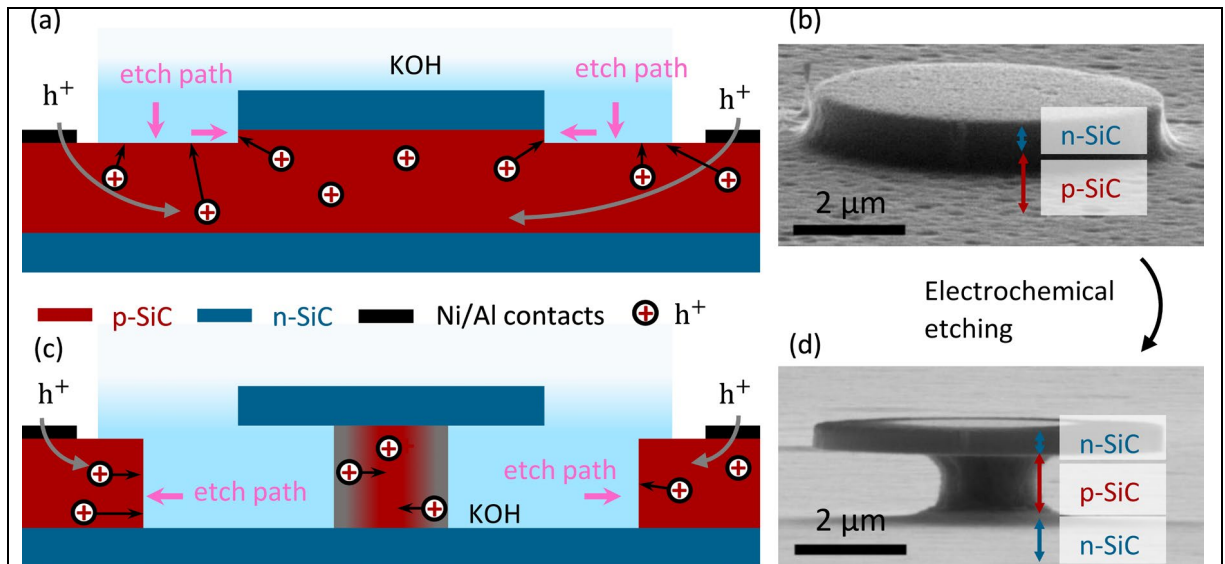

**Figure 2: Electrochemical etching using dopant-defined layers.** (a) and (b) before ECE, areas to be removed are defined by lithography and gas etching, slightly into the p-layer. Positive charge carriers (h+), required for ECE, are supplied via Ni/Al ohmic contacts. Applying anodic voltages results in (energy) band bending of the semiconductor, holes accumulate at the p-SiC-KOH interface and promote etching. (c) and (d) ECE removes the p-SiC layer. The etching is stopped vertically by the n-SiC layer; the lateral etching is stopped as soon as the remaining p-island is electrically unconnected. Without any applied potential, the p-SiC-KOH interface depletes of holes (due to band bending, grey area). (b) and (d) SEM micrographs, scale-bar: 2 µm.

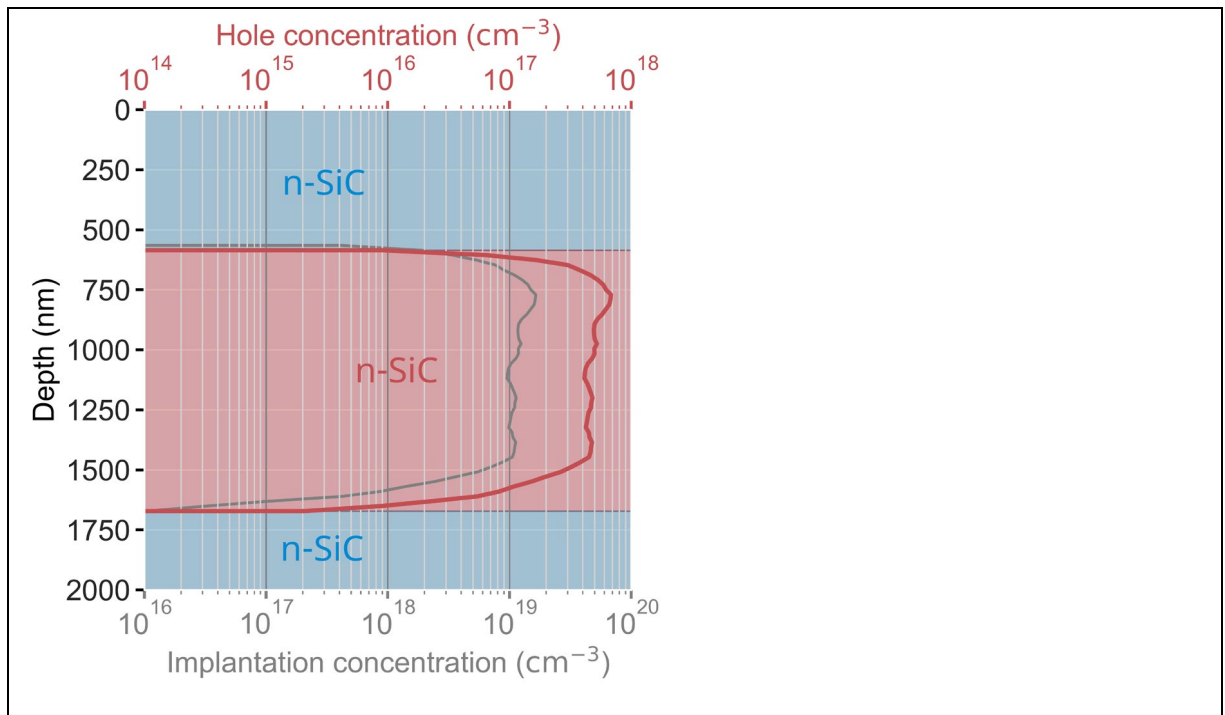

**Figure 3: Dopant-defined layers for electrochemical etching.** By suitable implantation profiles, a sharply defined p-SiC layer with an excess of positive charge carriers (holes) is defined. The hole concentration is calculated, utilizing the charge neutrality equation, assuming a compensation ratio of 0.35 [13], and a doping concentration dependant ionization energy ($E_{\text{ion}}(N_{\text{Al}}) = 210$ meV $- 3 \cdot 10^{-8}$eVcm $\cdot N_{\text{Al}}^{1/3}$) [14, 15].

This provides advantages as compared to gas etching, because it is isotropic. SiC is nearly inert, and only few chemical reactions are possible. The prerequisite for etching SiC is the presence of positive charge carriers (holes, $h^+$) at the surface. We opt for etching under alkaline conditions, the electrochemistry of which has extensively been studied by *van Dorp* [16, 17, 18]. In order to provide the required $h^+$ on the surface, a vast majority of publications uses electron-hole creation by ultraviolet light including our own work [19, 20, 5, 21]. This methodology is limited because of optical constraints, in particular it has poor vertical control [21]. Here we favor an electrochemical strategy where $h^+$-concentrations are created by appropriate doping patterns. Additional control can be gained by electric potentials. In a late stage of our investigations we found that the etching strategy is similar to [22].

The overall electrochemical reaction uses SiC and hydroxyl groups as educts, the products are gaseous CO and soluble silicate ions [17]:

$$SiC + 8OH^- + 6h^+ \rightarrow [Si(OH)_2]^{2-} + CO + 3H_2O.$$

This is a two-step process, where first in a solid state reaction silicon dioxide is formed ($SiC + 6OH^- + 6h^+ \rightarrow SiO_2 + CO + 3H_2O$) which is subsequently dissolved in KOH ($2OH^- + SiO_2 \rightarrow [Si(OH)_2O_2]^{2-}$).

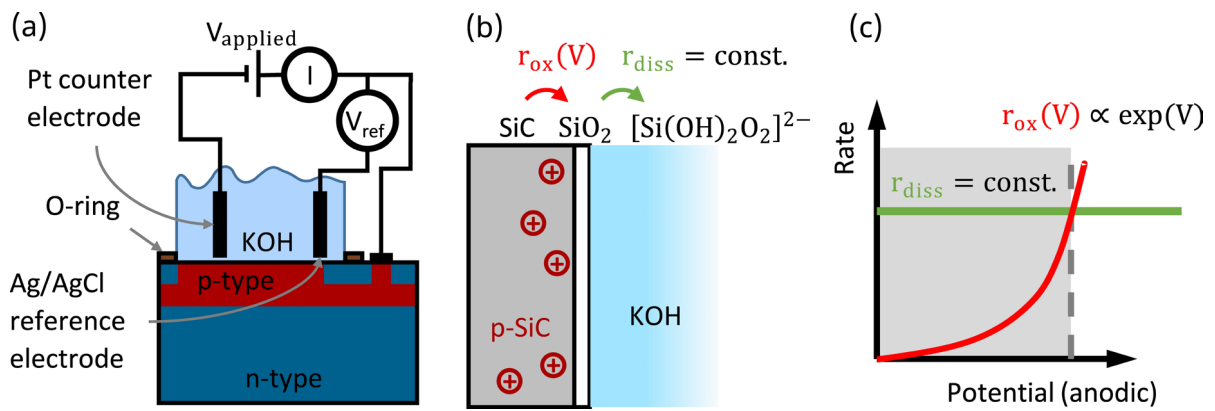

**Figure 4: Electrochemical oxidation reactions.** (a) Electrochemical etching setup. The current is entirely carried in the p-SiC layer. (b) Dissolution of SiC to $SiO_2$ and $SiO_2$ to a silicate product. (c) Oxidation ($r_{ox}$) and dissolution ($r_{diss}$) rates as a function of applied voltage. The grey area indicates the voltage range suitable for steady-state etching.

Note that both reaction rates have to be well balanced, because an overshooting of the $SiO_2$-formation would passivate the surface and stop the electrochemical process, see **Figure 4**.

This electrochemical reaction gives a handle to remove specific volumina selectively. p-doped, i.e. $h^+$-rich, regions can efficiently be etched as opposed to n-doped regions where etching is completely suppressed.

We start with a 4°-off-axis n-type wafer with an epilayer (nitrogen-doped $10^{16}$ cm$^{-3}$). The required doping profiles can be defined by ion implantation. An example hole concentration profile is shown in **Figure 3**, where Al-implantation creates a box-like p-type profile in a depth from 550 nm to 1.6 µm. At its flank, the hole concentration drops by more than ten orders of magnitude within 50 nm. In order to ensure reliable n-doping of the top layer, a counter-implantation with Nitrogen is performed. Subsequent annealing to 1700°C for 30 minutes in 900 mbar Argon-atmosphere re-establishes the crystalline lattice (locally, a carbon cap stabilizes the surface, [23]). Note that dopant diffusion is essentially absent in the rigid SiC-lattice.

For the geometries in this manuscript we used only vertical implantation profiles. The methodology can be readily extended to more complex 3D structures, when, in addition lateral patterning of the implantation is achieved, for example with robust metallic masks.

But also with laterally homogeneous doping profiles, 3D structures can be defined. For this purpose, we pattern resist masks which define a top window (electron beam lithography or similar). A projective etching of the n-type layer is performed by standard RIE / ICP-RIE techniques, such that the p-SiC layer is slightly etched. Now, the ECE is performed, which isotropically removes the p-SiC layer, see **Figure 2**. A typical lateral etch velocity is 2 µm/h. Care has to be taken that during this process an uninterrupted current path through the p-type layer has to be maintained. If however, during the etching p-type areas are disconnected from the current pathway, the etching stops for this island. This property can be exploited for the positive (see e.g. self-limited support columns for disk-shape optical resonators, see **Figure 2**). Maintaining intact current pathways thought the etching process has the rank of a design principle.

**Figure 1** displays a cantilever-like structure after ECE. We report one complication that arises after the ECE. Underneath the top n-type layer, in barely accessible regions, often an undesired porous p-type structure remains (goat beard).

It reminds the formation of porous SiC in KOH [24]. It can reliably be removed by two simple techniques: either a subsequent isotropic dry etch with $CF_4$ at 190 mTorr that is suited for well accessible devices like cantilevers. Alternatively, high temperature annealing beyond 1000°C in 900 mbar Argon atmosphere removes this layer even in hardly accessible regions. It can be suspected that thermal oxidation has a similar effect [25, 26].

An obvious quality criterion for optical or mechanical devices is the surface roughness. In our devices the ECE process leaves the etched surface quite smooth. The top surface of the n-SiC layer is essentially unchanged (in our experiments, $rms_{top}$= 1.46 nm, see **Figure 5a**). For characterizing the bottom layer, we removed a single clamped cantilever with scotch tape and studied its surface with the AFM. The result is shown in **Figure 5b**, it yields $rms_{bottom}$ = 2.48 nm. Hence, both the unetched top and the freshly etched bottom surfaces have both low surface roughnesses.

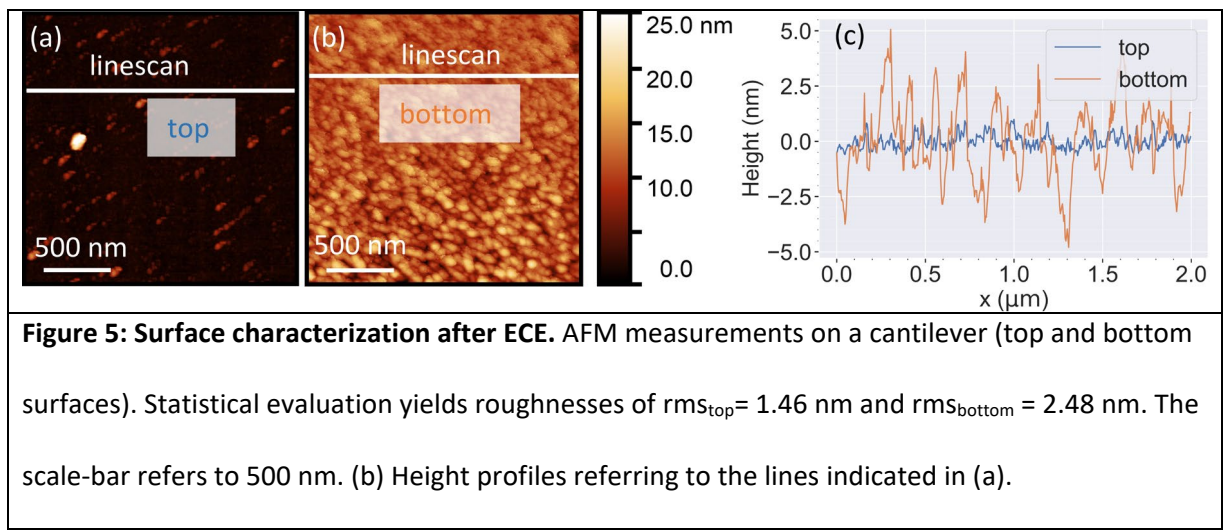

**Figure 5: Surface characterization after ECE.** AFM measurements on a cantilever (top and bottom surfaces). Statistical evaluation yields roughnesses of $rms_{top}$= 1.46 nm and $rms_{bottom}$ = 2.48 nm. The scale-bar refers to 500 nm. (b) Height profiles referring to the lines indicated in (a).

Finally, we address the question, how the presented 3D fabrication technique is compliant with further processing. Even underetched devices were spin-coated with PMMA and nLof resist materials without being damaged. They also survived lift-off processes, rinsing and drying without special precautions. Remarkably, the devices are also robust with respect to high temperature steps. In SiC, relevant spin carrying color centers are created, converted and finally annealed out in a temperature range from 400°C to 1400°C [12]. The native oxide layer sublimes at temperatures above 800°C in UHV [27]. Epitaxial graphene fabrication in n-type 4H-SiC is performed at 1400°C to 1500°C. Unintentional implantation damage anneals out at 1600°C to 1700°C. Hence, we explored this entire temperature range with completely processed devices like cantilever structures or membranes. They were exposed to high temperature steps in Ar atmosphere (900 mbar) for 30 minutes and subsequently investigated with SEM (see **Figure 6a - d**) and AFM (**Figure 6e**). The shape of the cantilevers is maintained at least up to 1550°C. Beyond this temperature, as can be seen in **Figure 6d** a visible re-arrangement occurs. It is most obvious at the lower edge, where a discontinuity occurs. Also in the lower and upper left corners, an additional faceted transition is formed, following crystalline directions. Remarkably, below 1550°C, our cantilevers provide an excellent shape stability. An analysis of the upper surface profile shows very little effect up to 1200°C. In the temperature range of 1275°C to 1350°C, pronounced terraces are formed and step-bunching occurs (4° miscut) with typical step heights of the order of 50 nm, beyond 1550°C approaching towards 100 nm.

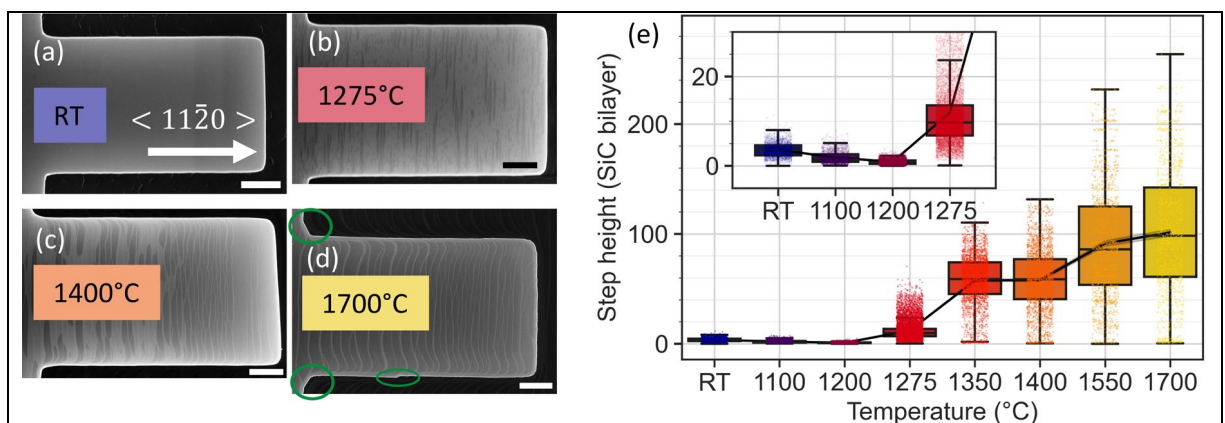

**Figure 6: Cantilever shape evolution during high temperature annealing.** (a) to (d) SEM micrographs of the very same cantilever after step-wise annealing at selected temperatures. Faceted transitions are highlighted in green. Scale-bar: 2 µm. (e) AFM line-scans were performed in the <11$\bar{2}$0> direction. The plot shows the analysis of the step height of this annealed cantilever

as a function of annealing temperature. The SiC step height significantly increases for annealing temperatures above 1275°C.

In the SEM micrographs, it becomes also apparent that epitaxial graphene starts to grow, which homogenously covers the surface and provides an inert and atomically smooth surface termination, as long as oxygen plasma is avoided.

In conclusion, we present a versatile electrochemical fabrication route for generating high-quality monolithic 3D devices in SiC, which paves the way to implement mechanically and optical devices on the SiC-platform, in addition to the already established electrical functionalities.

**Acknowledgements**

Support from Patrik Schmuki and Matthias A. Popp in early stages of this work is acknowledged. We further acknowledge financial support by German Research Foundation (DFG, QuCoLiMa, SFB/TRR 306, Project No. 429529648), project B03.

**Author contributions**

A.H. developed the etching strategy, fabricated samples and conducted the measurements. F.G. conducted the annealing experiments. M.N.M. fabricated the membrane. M.K. contributed his experience in SiC processing. H.B.W. conceived the experiment. The manuscript was written by A.H. and H.B.W. All authors discussed the results and contributed to the final manuscript.

**Competing interests**

The authors declare no competing interests.

**Data Availability Statement**

The data that support the findings of this study will be made available along with the publication in an open-access repository, with a suitable DOI.